# Single atom in a superoscillatory optical trap

Hamim Mahmud Rivy[1], Syed A. Aljunid[2], Emmanuel Lassalle[3+], Nikolay I. Zheludev[2,4], David Wilkowski[1,2,5,*]

[1]Centre for Quantum Technologies, National University of Singapore, 117543 Singapore, Singapore.

[2]Centre for Disruptive Photonic Technologies, The Photonics Institute, Nanyang Technological University, Singapore 637371, Singapore.

[3]Aix Marseille Université, CNRS, Centrale Marseille, Institut Fresnel, F-13013 Marseille, France

[4]Optoelectronics Research Centre, University of Southampton, Southampton, United Kingdom.

[5]MajuLab, International Joint Research Unit IRL 3654, CNRS, Université Côte d'Azur, Sorbonne Université, National University of Singapore, Nanyang Technological University, Singapore, Singapore.

*Corresponding author: david.wilkowski@ntu.edu.sg
+Current address: Institute of Materials Research and Engineering (IMRE), Agency for Science, Technology and Research (A*STAR), 2 Fusionopolis Way, Innovis #08-03, Singapore 138634, Republic of Singapore

## Abstract
Optical tweezers have become essential tools to manipulate atoms or molecules at a single particle level. However, using standard diffracted-limited optical systems, the transverse size of the trap is lower bounded by the optical wavelength, limiting the application range of optical tweezers. Here we report trapping of single ultracold atom in an optical trap that can be continuously tuned from a standard Airy focus to a subwavelength hotspot smaller than the usual Abbe's diffraction limit. The hotspot was generated using the effect of superoscillations, by the precise interference of multiple free-space coherent waves. We argue that superoscillatory trapping and continuous potential tuning offer not only a way to generate compact and tenable ensembles of trapped atoms for quantum simulators but will also be useful in single molecule quantum chemistry and the study of cooperative atom-photon interaction within subwavelength arrays of quantum emitters.

## INTRODUCTION
Optical trap is a pivotal instrument of quantum physics that allows the creation of one, two and three dimensional arrays of ultracold atoms [1], [2] or molecules [3], to explore cooperative light-matter interactions [4]–[8], topological quantum optics phenomena [9], [10], optical clocks [11], [12], quantum chemistry at a single molecule level [13], [14], quantum simulators [15], [16] and computers with Rydberg atomic states [17]–[20].

The performances of an optical trap depend on the ability to reduce its transverse size, such that one can optimally control the relative position of the atom and eventually get an inter-traps distance as small and precise as possible. The size of a conventional diffraction-limited trap can be defined by the Abbe radius of its optical hotspot $r_A = \lambda/2\text{NA}$, where $\lambda$ is the laser wavelength and NA is the numerical aperture of the lens. However, it is possible to reduce the size of the hotspot beyond Abbe's limit using the phenomenon of superoscillations that allows band-limited function to locally oscillate faster than its highest Fourier component [21]. Superoscillation fields can show subwavelength spatial variations of the intensity and phase of light, and also contain phase singularities and zones of energy backflow [22], [23]. Recently, superoscillation phenomena were used in super-resolution imaging and optical metrology [24], [25]. Here we report optical trapping of a single atom in a superoscillatory optical trap where the transverse size is subwavelength and below the Abbe's limit by a factor 0.69(3).

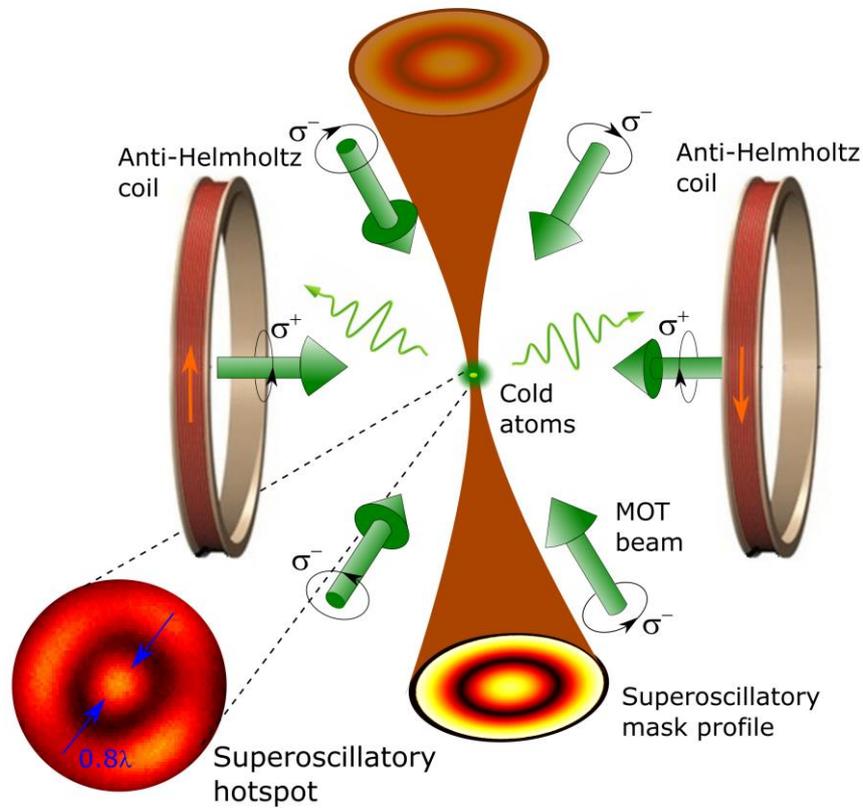

*Figure 1: **A superoscillatory trap for a single cold atom**. The trap is a combination of a magneto-optical trap consisting of a pair of anti-Helmholtz coils and six circularly polarized beams red-detuned with respect to the Cs $D_2$-line at 852 nm (green arrows) and a superoscillatory trap beam at the wavelength $\lambda = 1064$ nm having a subwavelength hotspot located at the position of the cloud of cold atoms. The superoscillatory mask profile is generated by spatial light modulators, transported to the pupil entrance of a long focal distance microscope objective located outside the vacuum chamber (see Fig.2a) and focused on the cloud of cold atoms. The image of the superoscillatory hotspot is obtained using a $75\times$ imaging system (see on Fig.2a). The 852 nm fluorescence photons (wiggling arrows) are collected on a photodetector (see on Fig.2a) and used to analyse the performances of the trap.*

## RESULTS

Initially, Cs atoms are cooled and trapped in a magneto-optical trap (MOT) operating on the $D_2$-line, at $\lambda = 852$ nm, as sketched in Fig. 1. The MOT consists of three orthogonal pairs of counter-propagating red-detuned laser (Green arrow) crossing at the null position of a magnetic-field quadrupole. The combined effect of resonant light scattering and magnetic field gradient leads to cooling and trapping of the atomic gas to the centre of the MOT (see Methods).

An optical trap made of a focused far-off-resonant laser beam at $\lambda = 1064$ nm is located at the center of the MOT (brown focused beam in Fig. 1). The transverse profile of the trapping beam is shaped thanks to reconfigurable amplitude and phase masks, generated by a pair of conjugated spatial light modulators (SLMs) [26], see Fig. 2a. A long working distance optical microscope (focal: 20.4mm) with a numerical aperture $NA = 0.43$ focuses the beam into the vacuum chamber hosting the MOT (see Supplementary Note 1). The SLMs, together with this focusing lens allow a continuous transformation of the trap potential from a conventional Airy profile to a narrower superoscillatory hotspot. A second optical microscope, confocal with the focusing lens, and a third lens form an $75\times$ magnification optical system used to image the profile into a CCD camera. Indeed, as the superoscillatory hotspot is formed by free space waves, it can be imaged by a conventional lens without loss of resolution. Two profiles, whose trapping performances will be later compared, are shown on Fig. 2b&c. The superoscillatory profile is characterized by a subwavelength hotspot with an Abbe radius of $0.85(3)\ \mu m = 0.80(3)\ \lambda = 0.69(3)\ r_A$, namely below the Abbe's limit (see Methods). In contrast, the Airy hotspot has a radius of $1.34(3)\ \mu m = 1.26(3)\ \lambda = 1.09(3)\ r_A$, just above the Abbe's limit, indicating a good control of the optical-system aberrations. The superoscillatory profile is characterized by a subwavelength hotspot and a more visible diffraction ring than usually obtained with an Airy profile [26]. The Airy-trap power is 23 mW. Because of the amplitude mask, the total power of the superoscillatory field is reduced to 8 mW. We measure 1.1 mW within the hotspot, corresponding to 14% of the total power. The remaining optical power is spread into the outer rings.

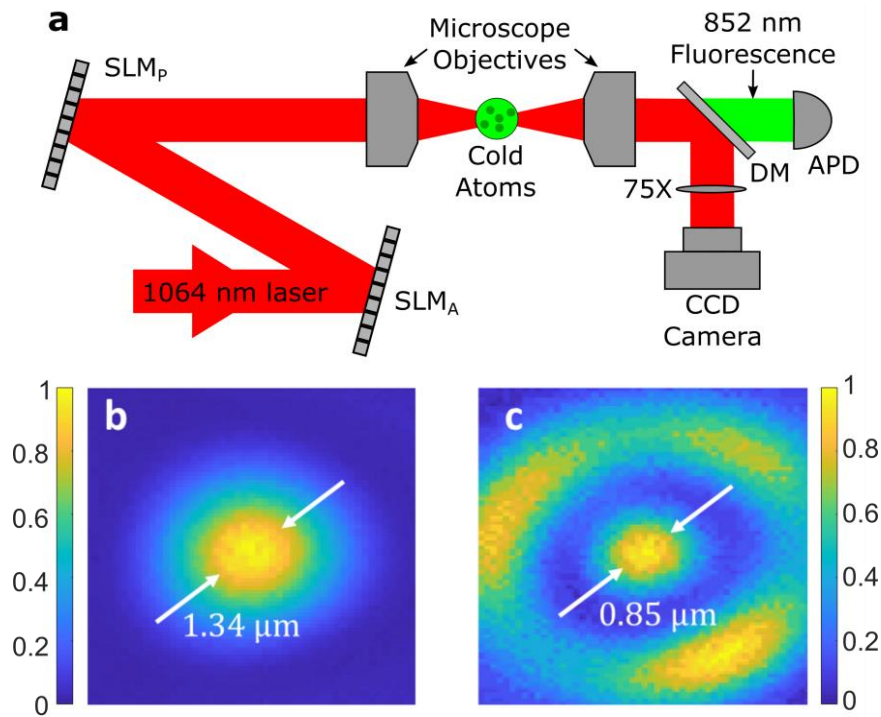

*Figure 2: **Generation of the superoscillatory optical trap. a** Schematic of the optical setup. A 1064 nm laser is focused at the cold-atoms cloud position using a microscope objective. The amplitude and the phase transverse profiles are shaped by two optically conjugated spatial light modulators, labelled $SLM_A$ and $SLM_P$, respectively. The beam profile is imaged on a CCD camera using a $75\times$ magnification optical system made of a second microscope objective and a lens. A dichroic mirror (DM) is used to separate the 1064 nm beam from the atomic fluorescence. The latter is sent to an Avalanche Photodiode (APD) in a photon counting mode. **b** Normalized image and size of the airy profile. **c** Normalized image of the superoscillatory profile, and size of the hotspot.*

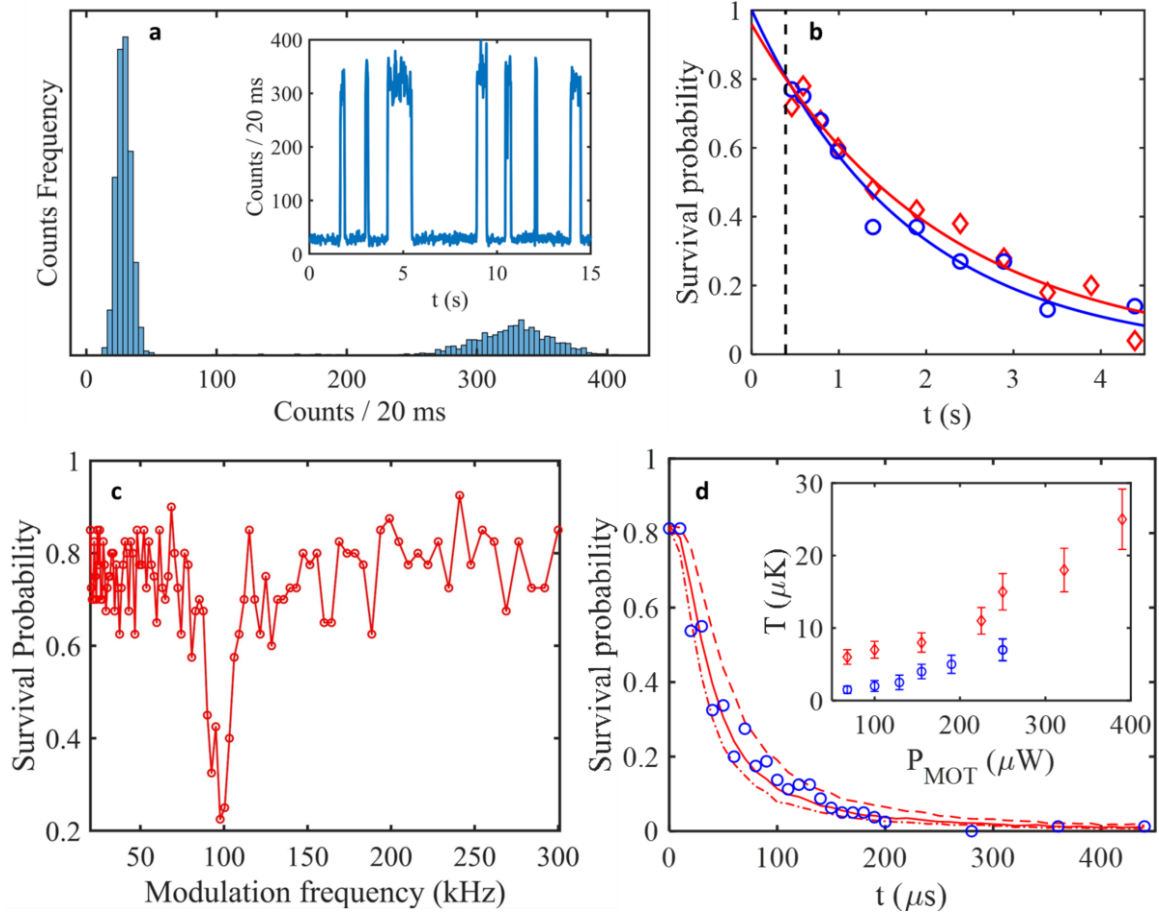

Figure 3: **Dynamics of a single atom in a superoscillatory optical trap. a** A histogram and a time dependence signal (inset) of the fluorescent photons integrated in 20 ms time intervals. **b** The survival probability of an atom in the optical trap for superoscillatory (blue circles) with a hotspot power of 1.1 mW, and Airy profile (red diamonds) with a hotspot power of 3 mW as function of the switch-off time of the MOT beams. The blue and the red curve are exponential decay fits of the superoscillatory and Airy trap with similar trap depth, respectively. The black vertical dashed line at $t = 0.39\ s$ indicates the total time of the transfer sequence from the Airy to superoscillatory profiles and vice-versa (see Supplementary Note 2). **c** The survival probability of an atom in the superoscillatory optical trap versus the modulation frequency of the trap beam power indicating a trapping frequency of $\sim 50\ kHz$. Outside the resonance, the survival probability reads 0.76(7), where the error corresponds to one standard deviation. **d** The survival probability of an atom in the superoscillatory trap after switching off the trap beam for a time t (blue circles). The dashed, plain and dashed-dotted red curves show results of the survival probability modelled for an effective atomic temperature of $1.5\ \mu K$, $2.5\ \mu K$, and $3.5\ \mu K$ respectively. We then deduce an effective temperature of 2.5(10) $\mu K$ here, for an optical power per MOT beam of $130\ \mu W$. The inset shows the effective temperature of the atom in the superoscillatory (blue circles) and Airy hotspot (red diamond) as functions of optical power per MOT beam. The error bars represent one standard deviation of the mean. In panels b, c, and d, each data point corresponds to the averaged survival probability over 80 runs. The standard deviation is then estimate to be $\sim 0.1$.

Characterisations and diagnostics of the population in the optical trap are undertaken via the dynamics of the fluorescence emission at 852nm, detected by an avalanche photodiode through the second optical microscope and a dichroic mirror to reject the 1064 nm field (see Fig. 2a and Supplementary Note 1). In the inset of Fig. 3a, we show a typical time-dependence of the fluorescence recorded for the Airy hotspot profile indicating the dynamics of the atomic population in the trap, with two well-separated plateaus. It leads to a histogram of the count occurrence with two distinct distributions at a mean count of 30 and 330 per 20 ms integration time corresponding to the noise level of the detector and the presence of a single atom in the trap with a probability of ~30%, respectively (see Fig. 3a). We note that larger atom numbers in the optical trap are suppressed by collisional blockade mechanisms [27], [28].

We did not observe direct loading of the superoscillatory pattern, including the hotspot and the outer ring from the cold atom cloud. This is likely due to the small trap volume of the hotspot and the weak trapping depth of the outer ring. An abrupt reconfiguration of the trap from the Airy to superoscillatory profile also rarely results in keeping the trapped atom. However, a multi-step quasi-adiabatic reconfiguration of the initial Airy trap allows a transfer to the superoscillatory hotspot. In practice, using two intermediate transverse profiles is enough to obtain a transfer from Airy to superoscillatory trap with a probability close to unity. Since the Airy profile does not overlap with the superoscillatory outer rings, we can safely rule out transfer and trapping outside the superoscillatory hotspot. Measured by switching off the MOT, the lifetime of the atom in the superoscillatory trap is characterized by a 1/e decay time of $\tau = 1.8(3)$ s (Fig. 3b, blue open circle). With the Airy profile, the lifetime of the atom in the hotspot has a similar 1/e decay time of $\tau = 2.1(3)$ s (red open diamonds). In addition, we found that in the Airy profile configuration the lifetime weakly depends on the trapping laser power indicating that the loss mechanism is likely dominated by the collision with the thermal background gas. We note that the adiabatic transfer from Airy to superoscillatory profile takes 300 ms, whereas the transfer from superoscillatory to Airy trap takes 90 ms. Taking the total transfer time of 390 ms into account in the lifetime, the 1/e decay extrapolation at the time origin shows a survival probability close to unity for both Airy and superoscillatory hotspot. It indicates that no extra significant losses occur during the transfer from Airy to superoscillatory trap.

We evaluate the transverse trapping frequency by modulating the light power of the trapping beam. If the frequency of modulation is twice the trapping frequency, the atom gains kinetic energy parametrically, shortening its survival probability. Fig. 3c show a dip of the survival probability at ~ 100 kHz, thus corresponding to a trapping frequency of the superoscillatory trap of $f_{SH} \sim 50$ kHz. A measurement on the Airy hotspot gives a trapping frequency of $f_{AH} \sim 75$ kHz. We note that the trapping frequency scales as $f \sim \sqrt{P}/\sigma$, where $P$ is the optical power and $\sigma$ is trap characteristic transverse cross-section. Thus, we expect $f_{AH}/f_{SH} \sim 1.8$, in reasonable agreement with the measurements (see Methods).

The survival probability measured when switching off for a short time the superoscillatory trap laser is presented in Fig. 3d. To evaluate the effective temperature, we matched the survival-probability dependence with a simulation assuming a Maxwell-Boltzmann distribution of atomic velocities. The inset shows the effective temperature in the Airy and superoscillatory traps as a function of the MOT laser power. At minimal powers, it reaches $6(1)$ μK with Airy profile and $1.5(5)$ μK with superoscillatory profile. The temperature reduction observed in the superoscillatory hotspot is likely due to adiabatic cooling when transferring the atom from the Airy to the superoscillatory trap [29] (see Methods).

## CONCLUSION

In summary, we reported the first realization of a subwavelength single ultracold atom trap with a superoscillatory optical field, which is the smallest trap in the linear regime demonstrated so far. The atom reaches an effective temperature of about 1.5 $\mu K$ and a lifetime of about 1.8 seconds, limited by the residual background pressure.

Our superoscillatory hotspot reduces the area of transverse localization of the atom by a factor of 2.5 in comparison with an Airy profile trap for the same bandwidth of the optical system. Further reduction of the superoscillatory hotspot size could be realized using more detailed engineering of the field transverse profile [30] and larger optical power. In addition, nonlinear response on a three-level atomic scheme has been either explored to create subwavelength lattice traps [31], and subwavelength lattice spacing [32] or proposed for optical nanotweezers [33]. Similar approaches could be also implemented using a superoscillatory optical trap to further reduce the trap size.

We measured a transverse trapping frequency of $\sim 50$ kHz for a superoscillatory hotspot with 1.1 mW power. However, since a larger power is available for the Airy hotspot, the trapping frequency remains larger for that configuration, leading to an overall better confinement. Nevertheless, we argue that in many applications the trapping-laser peak intensity is the limiting factor, for example to avoid spontaneous photon emission or multi-photonic absorption induced by the trapping field. In those cases, superoscillatory traps have a clear advantage over the Airy profile traps as they allow smaller trap sizes thus, similar trapping frequency at lower laser intensity. Superoscillatory traps can find applications when compact and tuneable ensembles of trapped atoms are needed for quantum simulation of many-body effects, cooperative quantum metasurfaces, and for single molecule quantum chemistry.

Creation of standard optical trap arrays with Gaussian beams is facilitated by the absence of outer ring [1]. With superoscillatory field, the presence of the optical field outside the hotspot is unavoidable. So far, superoscillatory fields are generated with super-resolution imaging applications in mind, meaning that the figure of merit, besides having the smallest spot, consists in pushing the outer ring as far as possible from the hotspot, regardless of the resulting spatial distribution of the field [34]. For a hotspot array, one might need a new paradigm where the optical field outside the hotspot shall be spatially spread to reduce the peak energy density in outer rings. For practical realisation of a hotspot array, one can envision to create an optical trap array with Gaussian beams, which can be transferred in a hotspot array as discussed in this communication. The latter will be adiabatically compressed to form a sub-wavelength array. To avoid spurious interference, one could consider alternatively turning on neighbouring sites, in the same spirit as Ref. [16].

## METHODS

**Magneto-optical Trap (MOT)**: The Cs MOT is generated using three quasi-orthogonal retro-reflected optical beams. The beams are detuned by $-4\Gamma$ from the $6S_{1/2}$ ($F = 4$) $\rightarrow 6P_{3/2}$ ($F = 5$) D$_2$-line, where $\Gamma/2\pi = 5.2$ MHz is the transition linewidth. The usual optical power per beam is 130 µW with a beam waist of $\sim$750 µm, leading to an on-resonance free-space saturation parameter of $s_0 \sim 13$. A repumper beam (total optical power 3 mW), tuned on resonance with the $6S_{1/2}$ ($F = 3$) $\rightarrow 6P_{3/2}$ ($F = 4$) transition, is overlapped with the cooling beams. A pair of coils in an anti-Helmholtz configuration generates a magnetic field gradient of 3.5 G/cm. The thermal Cs vapour is obtained from a dispenser that we occasionally operated at 2.5 A to maintain the background Cs vapour at a low pressure.

**Determination of the optical trap profile**: The Airy and superoscillatory profiles are collected on a CCD camera through a 75× imaging system (see Fig. 2a). The images are fitted with an Airy disk pattern, to extract the Abbe radius. For a diffraction-limited spot, the Abbe radius is $r_A = \lambda/2\text{NA} = 1.15\,\lambda = 1.24\,\mu m$, with $\lambda = 1064\,nm$. For the Airy profile, we find $1.34\,\mu m = 1.09 r_A$, which is very close to the diffraction limit. For superoscillatory field, we get a subwavelength spot size of $0.85\,\mu m = 0.69\,r_A$, well below the Abbe's limit. See Supplementary Note 1 for more details.

**Determination of the temperature**: The effective temperature of the atom in the optical trap is estimated using a release and recapture technique [29]. It consists of measuring the survival probability after switching off the optical trap for a variable time. We compare the experimental data with a release and trap simulation where the trap has an Airy disk shape with a radial form $U(r) = -U_0[2r_{Am}J_1(\pi r/r_{Am})/\pi r]^2$, where $J_1(x)$ is the Bessel function of the first-kind, $r_{Am}$ is the measured Abbe radius and $U_0 = 500 k_B$ µK is the expected trap depth for a power of 1.1 mW in the superoscillatory hotspot, $k_B$ is the Boltzmann constant. Since the superoscillatory tri-dimensional (3D) profile has a needle shape along the laser propagation axis [34], we assume that the atom can be lost only through its dynamic into the transverse plane. Here, we reduce the general 3D problem into a 1D problem, considering the transverse radial direction of the trap. Then, we perform a statistically relevant number of trials, and we consider that the atom has left the trap either if $vt \geq r_1$ ($r_1 \approx 1.22\, r_{Am}$ is the radius of the first zero of the Bessel function) or if the classical mechanical energy of the atom after switch on the trap is positive, namely $mv^2/2 + U(vt) > 0$, where $t$ is the switch-off time and $v$ is the initial radial velocity. The Cartesian transverse components of velocity $v_i$ ($i = x, y$) are random Gaussian variables with $\langle v_i \rangle = 0$, and $\langle v_i^2 \rangle = k_B T/m$, where $T$ and $m$ are the temperature and the atom's mass, respectively. Since $T \ll k_B U_0$, we disregard the initial spatial dispersion of the atom.

**Adiabatic cooling**: We transfer in a quasi-adiabatic manner the atom from the Airy to the superoscillatory hotspot, and we observe a reduction of the temperature by a factor ranging from two to three (see inset of Fig. 3d). Moreover, it was shown that the reduction of temperature during adiabatic cooling is proportional to the reduction of the trapping frequency [29]. In our system, we estimate the reduction of the trapping frequency as such. During the transfer, the optical trap transverse size is reduced by a factor of ~1.6, whereas the power is reduced by a factor ~21, with 23 mW in the Airy hotspot and 1.1 mW in the superoscillatory hotspot. Hence, while transferring the atom from Airy to superoscillatory profile, the transverse trapping frequency is expected to reduce by a factor ~$\sqrt{21}/1.6^2 \approx 1.8$, whereas a direct measurement of the trapping frequency using the modulation of the trapping beam amplitude technique gives a reduction of ~1.5 in reasonable agreement with the expected value. Longitudinally, we expect a more pronounced reduction of the trapping frequency, because of the needle shape of the superoscillatory hotspot [34]. Even if we do not have a precise idea of the reduction of the longitudinal trapping frequency and its impact on adiabatic cooling, we note that the reduction of the averaged trapping frequency is in the same order of magnitude as the temperature reduction. It confirms the dominant role of adiabatic cooling during the transfer from the Airy to the superoscillatory hotspot.

## ACKNOWLEDGEMENTS
The authors thank Eng Aik Chan for his early contribution to the experimental setup and Edward T. F. Rogers for his help on implementing the superoscillation optical system. DW thanks Steven Touzard for fruitful discussions. EL thank Institut Fresnel (via Fonds pour la Science 2018) and Nanyang Technological University for supporting his stay in Singapore. This work was supported by the Centre for Quantum Technologies funding Grant No. R-710-002-016271, the Singapore Ministry of Education Academic Research Fund Tier3 Grant No. MOE2016-T3-1006(S), the Singapore Ministry of Education Academic Research Fund Tier 1 Grant No. RG160/19(S) and the Engineering and Physical Sciences Research Council, UK (grant numbers  EP/T02643X/1)

## AUTHORS CONTRIBUTIONS
DW and NIZ conceived the idea of superoscillatory atomic trap and supervised the work. HR, EL and SA built up the experiment and collected the experimental data. HR, SA, and DW analysed the data. All authors contributed to the writing of the manuscript.

**Supplementary Note 1: Superoscillatory fields optical set-up**

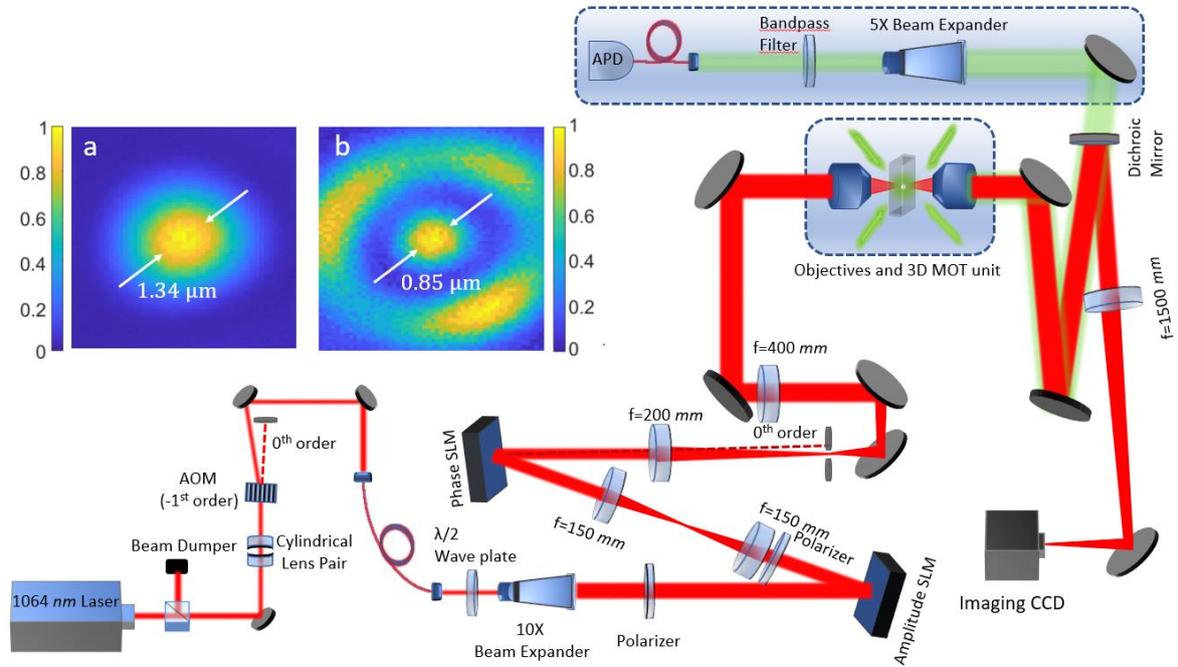

*Supplementary Figure 3. **The optical setup, amplitude and phase masks**. Main picture: Sketch of the OT optical system. The phase and amplitude profiles of a 1064 nm beam (red optical path) are engineered thanks to two conjugated SLMs. The beam is focused at the centre of a cold atoms cloud in a Magneto-Optical Trap (MOT) represented here by laser beams (green arrows). The beam profile is imaged on a CCD camera with a 75x magnification optical system. The cold atoms fluorescence (green optical path), due to the MOT beams is collected on an APD. (a) and (b) Images of the Airy and the SO profiles, with an Abbe radii of 1.34(3) μm and 0.85(3) μm, respectively.*

The optical trap (OT) is created thanks to a 1064 nm linearly polarised 1.5 W laser. The OT beam is spatially filtered by a single mode fibre and expanded to a beam waist of ∼10 mm (see Supplementary Fig. 1). The core of the superoscillatory (SO) optical setup consists of two optically conjugated spatial light modulators (SLMs) for transversally resolved amplitude and phase reflection. The SLMs system is set to generate an SO optical field profile constructed from two circular prolate spheroidal functions $E(r/\lambda) = 3.123 S_2(r/\lambda) + S_3(r/\lambda)$, where r is radial distance from the hotspot centre, as described in [1].

The active aperture diameter of our SLM-based optical system is 9.65 mm. The SLMs are imaged at the entrance pupil of a high numerical aperture microscope objective (working distance of 20.2 mm, effective focal length 20.4 mm), through a 2× magnification achromatic lens doublet. The entrance pupil diameter of the microscope is 20.4 mm, therefore slightly larger than the diameter $2 \times 9.65 = 19.3$ mm of the SLMs image, leading to a numerical aperture of NA = 0.43.

We use two microscope objectives, customized to compensate for the spherical aberration of the vacuum chamber windows, and provide a diffraction limited optical system at 852 nm and 1064 nm. The first objective focuses the 1064 nm OT beam at the cold atoms cloud position. The second objective collects the fluorescence signal at 852 nm that is sent to an avalanche photodiode detector (APD), and serves to image the 1064 nm OT on a CCD camera. The latter imaging system is made of the second objective and a $f = 1500$ mm lens, and provide an image of OT beam with a 75× magnification. The images of the Airy profile used to load the OT and the SO profile are shown in Supplementary Fig. 1a and Fig. 1b, respectively.

**Supplementary Note 2: Temporal sequence for superoscillation trapping characterizations**

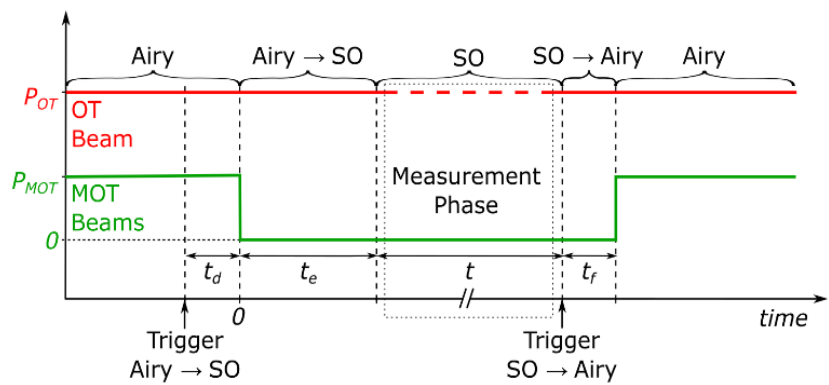

*Supplementary Figure 2.* **Temporal sequence for superoscillation trap characterizations**. *The different sequences of the OT beam profiles are given on the top of the figure. The red and the green curve indicate the temporal evolution of the OT beam power and the MOT beams (cooling and repump beams) power, respectively. The measurement is performed during the SO phase (dotted rectangle). Here, the OT beam power might change (dashed red line) according to the type of measurement (see text for more details). At time $-t_d$, the Airy trap is loaded with a single atom and a trigger is sent to start the transfer from Airy to SO profile. $t_d = 140\ ms$ is the delay time associated with the software trigger, $t_e = 300\ ms$ is the time to change form Airy to SO profile, and $t_f = 90\ ms$ is the time to transfer back to Airy OT for measurement of the survival probability.*

The atom population in the Airy OT is determined from the APD photo-count integrated over 20 ms, see Fig. 2a in the main text. If the APD photo-count is above 100, the OT is considered to be populated with one atom, which triggers the temporal sequence sketched in Supplementary Fig. 2. During $t_d = 140$ ms, the MOT beams power is eventually either decreased or increase (not shown in Supplementary Fig. 2), to change the effective temperature of the atom in the OT. At $t = 0$, the MOT and repump beams are switched off and we adiabatically change the Airy profile into a SO profile during $t_e = 300$ ms. The SO OT is maintained for an extra variable time $t$ to perform different type of measurements. Then, we send a trigger to go back to Airy profile within the next $t_f = 90$ ms. Finally, the MOT beams are turned on to measure the survival probability of the atom in the OT. Overall, the MOT beams are switched off during a time of $t + t_e + t_f = t + 390\ ms$.

During the measurement phases, we act on the OT power, according to the type of measurement, we are aiming to perform. As discussed in the main text, we did three measurements:

- The lifetime of the atom in the OT, which is done while maintaining the OT power at constant value.
- The trap frequency measurement, which is performed by modulating the OT beam during 300 ms with 4% of modulation depth. For each run, the modulation frequency is set within 10 kHz to 300 kHz.
- The effective temperature measurement, which is done by turning off the OT beam for a variable time $t = 0 - 400$ μs

We performed 80 runs for each data point shown in Supplementary Fig. 2b-d in the main text. To avoid possible systematic drifts, we randomized the order of the parameter-value occurrence.